# Classification of Cervical Cancer Dataset

**Abstract ID:** 2423

**Y. M. S. Al-Wesabi, Avishek Choudhury, Daehan Won**
**Binghamton University, USA**

## Abstract

Cervical cancer is the leading gynecological malignancy worldwide. This paper presents diverse classification techniques and shows the advantage of feature selection approaches to the best predicting of cervical cancer disease. There are thirty-two attributes with eight hundred and fifty-eight samples. Besides, this data suffers of missing values and imbalance data. Therefore, over-sampling, under-sampling and imbedded over and under sampling have been used. Furthermore, dimensionality reduction techniques are required for improving the accuracy of the classifier. Therefore, feature selection methods have been studied as they divided into two distinct categories, filters and wrappers. The results show that age, first sexual intercourse, number of pregnancies, smokes, hormonal contraceptives and STDs:genital herpes are the main predictive features with high accuracy with 97.5%. Decision Tree classifier is shown to be advantageous in handling classification assignment with excellent performance.

## 1. Introduction
Cervical cancer is the most common cancer among women in developing countries, the WHO report [1]. In the United States, there are 129,001 new cases in 2015 despite the provided healthcare facilities, where 273,000 deaths in 2002 worldwide [1]. Cervical cancer dataset has been published in 2017 by [2], which involves 858 samples and 32 features as well as four targets. These attributes include demographic information, habits like smoking and historic medical records. The complexity of this data is the multiple screening and diagnosis approaches that leads to a complex ecosystem. Consequently, the prediction of patient's factor risk and the best screening strategy is a main problem.

Cervical cancer data has been studied from different researchers in last few years . [2] is the first contributor such that transform learning method was the main purpose of the study to show its impact on sharping the accuracy. [3] have studied the same data using cost-sensitive classification with Decision Tree classifier. The good results have been reported by [4] such that Support vector machine with different approaches were applied to yield almost 94% accuracy. However, cervical cancer disease has been deliberated from different angles. For instance, [5] pointed out the cervical cancer control in HIV-positive women and expectation of reducing the mortality in 2030. Interesting investigation has been conducted by [6] to develop novel biomarkers for early diagnosis from specific proteins, enzymes and metabolites.

There are three main steps in the data mining, preprocessing, classification process and the decision-making with analysis [3]. This data includes 3,622 missing values out of 27,456 observations, which forms 13.2% of the data. This survey investigates different machine learning classifiers such as Gaussian Naive Bayes (GNB), Decision Tree (DT), Logistic Regression (LR), k-nearest neighbors (KNN) and Support Vector Machines (SVM), see [7-12]. Further investigation involves wrapper methods such as Sequential Feature Selector, both Forward and Backward version. Some recommended resources for feature selection techniques are available in [13-16].

## 2. Data Description

Cervical cancer data involves 858 samples and 32 features as well as four classes (Hinselmann, Schiller, Cytology and Biopsy) has been published in [2]. This paper focuses on studying the Biopsy target as it recommended by the literature review.

Table 1: Attributes and their types

| Attribute | Type | Attribute | Type | Attribute | Type |
|---|---|---|---|---|---|
| Age | Integer | STDs | Bool | STDs:HIV | Bool |
| Number of sexual partners | Integer | STDs (number) | Integer | STDs:Hepatitis B | Bool |
| First sexual intercourse (age) | Integer | STDs:condylomatosis | Bool | STDs:HPV | Bool |
| Number of pregnancies | Integer | STDs:cervical condylomatosis | Bool | STDs: Number of diagnosis | Integer |
| Smokes | Bool | STDs:vaginal condylomatosis | Bool | STDs: Time since first diagnosis | Integer |
| Smokes (years) | Bool | STDs:vulvo-perineal condylomatosis | Bool | STDs: Time since last diagnosis | Integer |
| Smokes (packs/year) | Bool | STDs:syphilis | Bool | Dx:Cancer | Bool |
| Hormonal Contraceptives | Bool | STDs:pelvic inflammatory disease | Bool | Dx:CIN | Bool |
| Hormonal Contraceptives (years) | Integer | STDs:genital herpes | Bool | Dx:HPV | Bool |
| IUD | Bool | STDs:molluscum contagiosum | Bool | Dx | Bool |
| IUD (years) | Integer | STDs:AIDS | Bool | | |

## 3. Methodology

This section concentrates on the methodology of this survey, which can be described into three main parts. First, the preprocessing experiments, this involves missing values treatment using standard measurements like the mean for numerical values and mode for categorical attributes. Several patients decided to not answer some questions due to personality. As a result, 13% of total questions were missed. There are two features with 92% of missing values which are STDs: Time since first diagnosis and STDs: Time since last diagnosis, so they have been omitted. Secondly, five classifiers including the GNB, KNN, DT, LR and SVM are applied to figure out the appropriate classifier for this data. Imbalanced data have been studied by applying three techniques, over-sampling, under sampling and both together. Eventually, for sharping the results and looking at the main risk factors of the cervical cancer, feature selection methods are applied like wrapper methods. Wrapper methods such as Sequential Feature Selector, both Forward and Backward version are used. Table 2 shows some basic notations.

Table 2: Basic notations

| Term | Formula | Definition |
|---|---|---|
| **Accuracy** | (TP + TN)/(P+N) | Rate of the correct prediction for both healthy and not healthy patients |
| **Sensitivity=recall= true positive rate** | TP/(TP+FN) | The percentage of sick people who are correctly identified as having the disease. |
| **Specificity= true negative rate** | TN/(FP+TN) | The percentage of healthy people who are correctly diagnosed as healthy. |
| **Precision** | TP/(TP+FP) | positive predictive value |
| **F-measure** | (2 x recall x precision ) / (recall+precision) | Harmonic mean that combines Precision and recall. |

# 4. Results

The data has been divided into 758 training and 100 samples are testing except some experiments that will be specified. Python is the used software for whole experiments.

## 4.1 The basic classification

The result shows that the LR, SVM and KNN perform better than DT and GNB. Besides, the GNB is the worst classifier for this data.

Table 3: Five classifiers result

| Classifier /Performance | Gaussian Naive Bayes | Decision Tree | Logistic regression | SVM | KNN |
|---|---|---|---|---|---|
| Accuracy | 5.696 | 90.116 | 93.671 | 93.671 | 93.671 |
| Sensitivity=TP | 100.0 | 96.226 | 95.484 | 95.484 | 95.484 |
| Specificity | 4.790 | 15.385 | 0.000 | 0.000 | 0.000 |
| Precision | 1.325 | 93.293 | 98.013 | 98.013 | 98.013 |
| F-measure | 2.614 | 94.737 | 96.732 | 96.732 | 96.732 |

Due to the imbalancy on our data set, we have this results. So, we need to resolve this problem.

## 4.2 Imbalanced data

To address the biased data, there are several methods available. This study will apply SMOTETomek (combine method), under-sampling and over-sampling method that available in *imblearn class* in Python language.

### 4.2.1 SMOTETomek method

Table 4: Five classifiers results using combine sampling method

| Classifier /Performance | Gaussian Naive Bayes | Decision Tree | Logistic regression | SVM | KNN |
|---|---|---|---|---|---|
| Accuracy | 50.633 | 91.772 | 72.152 | 85.759 | 84.494 |
| Sensitivity=TP | 100 | 91.464 | 67.961 | 93.985 | 96.694 |
| Specificity | 49.677 | 92.105 | 80.0 | 79.781 | 76.92 |
| Precision | 3.703 | 92.593 | 86.419 | 77.160 | 72.222 |
| F-measure | 7.142 | 92.025 | 76.087 | 84.746 | 82.686 |

The DT classifier performs the best in this experiment, while the GNB performs poorly. Therefore, the GNB will be omitted from further studies.

### 4.2.2 Under-sampling

Table 5: Four classifiers results using under-sampling method

| Classifier /Performance | Decision Tree | Logistic regression | SVM | KNN |
|---|---|---|---|---|
| Accuracy | 68.182 | 54.545 | 68.182 | 54.545 |
| Sensitivity=TP | 77.778 | 66.667 | 72.727 | 60.0 |
| Specificity | 61.538 | 50.0 | 63.636 | 50.0 |
| Precision | 58.333 | 33.333 | 66.667 | 50.0 |
| F-measure | 66.667 | 44.444 | 69.565 | 54.545 |

This is expected results due to the small samples size, it was reduced from 800+ to only 100+.

### 4.2.3 Oversampling

Table 6: Four classifiers results using over-sampling method

| Classifier /Performance | Decision Tree | Logistic regression | SVM | KNN |
|---|---|---|---|---|
| Accuracy | 93.788 | 63.975 | 81.677 | 90.683 |
| **Sensitivity=TP** | 100.0 | 62.32 | 79.569 | 100.0 |
| **Specificity** | 88.439 | 67.289 | 84.559 | 83.606 |
| **Precision** | 88.165 | 79.289 | 87.574 | 82.249 |
| **F-measure** | 93.711 | 69.792 | 83.380 | 90.259 |

Table 7: Comparison between the DT classifier and corresponding imbalanced sampling methods

| Classifier /Performance | Over-sampling | Under-Sampling | Both |
|---|---|---|---|
| **Accuracy** | 93.788 | 68.182 | 91.772 |
| **Sensitivity=TP** | 100.0 | 77.778 | 91.464 |
| **Specificity** | 88.439 | 61.538 | 92.105 |
| **Precision** | 88.165 | 58.333 | 92.593 |
| **F-measure** | 93.711 | 66.667 | 92.025 |

**4.3 Feature selection based on Sequential Backward Selection (SBS) method**
This experiment focuses on wrapper methods, practically SBS and SFS with 12-selected features and 10-cross-validation.

| Classifier | Indices |
|---|---|
| LR | 5, 6, 7, 8, 9, 11, 13, 15, 19, 22, 26, 27 |
| SVM | 0, 1, 2, 3, 5, 8, 9, 11, 14, 25, 27, 28 |
| DT | 0, 1, 2, 3, 7, 8, 14, 16, 17, 20, 21, 23 |
| KNN | 0, 1, 2, 5, 6, 7, 8, 11, 12, 25, 28, 29 |

Table 8: Four classifiers results using the SBS method

| Classifier /Performance | Decision Tree | Logistic regression | SVM | KNN |
|---|---|---|---|---|
| **Accuracy** | 95.652 | 64.596 | 90.062 | 90.062 |
| **Sensitivity=TP** | 100.0 | 62.790 | 91.017 | 100.0 |
| **Specificity** | 91.617 | 68.22 | 89.032 | 82.702 |
| **Precision** | 91.716 | 79.881 | 89.941 | 81.065 |
| **F-measure** | 95.679 | 70.313 | 90.476 | 89.542 |

**4.4 Feature selection based on Sequential Forward Selection (SFS) method**

| Classifier | Indices |
|---|---|
| LR | 2, 5, 7, 8, 10, 11, 14, 15, 17, 19, 26, 27 |
| SVM | 0, 1, 2, 3, 4, 5, 7, 8, 9, 11, 12, 29 |
| DT | 0, 1, 2, 3, 4, 7, 16, 18, 19, 23, 24, 27 |
| KNN | 0, 2, 3, 4, 7, 11, 14, 15, 19, 22, 26, 28 |

Table 9: Four classifiers results using the SFS method

| Classifier /Performance | Decision Tree | Logistic regression | SVM | KNN |
|---|---|---|---|---|
| **Accuracy** | 95.652 | 65.838 | 90.373 | 90.062 |
| **Sensitivity=TP** | 100.0 | 63.849 | 91.071 | 100.0 |
| **Specificity** | 91.617 | 69.725 | 89.61 | 82.703 |
| **Precision** | 91.716 | 80.473 | 90.532 | 81.065 |
| **F-measure** | 95.679 | 71.204 | 90.801 | 89.542 |

Both Forward and Backward feature selection techniques are comparable and equivalent.

**4.5 Selecting some features**
Another interesting result can be obtained by selecting the common factors between the DT and the KNN. Then by applying the main classifiers, it yields the following results.

Table 10: Four classifiers results using selective features

| Classifier /Performance | Decision Tree | Logistic regression | SVM | KNN |
|---|---|---|---|---|
| **Accuracy** | **97.515** | 49.068 | 88.198 | 87.577 |
| **Sensitivity=TP** | **100.0** | 51.533 | 90.683 | 100.0 |
| **Specificity** | **95.031** | 46.540 | 85.714 | 79.274 |
| **Precision** | **95.266** | 49.704 | 86.390 | 76.331 |
| **F-measure** | **97.576** | 50.602 | 88.485 | 86.577 |

Overall, it is noticeable that the DT is the recommended classifier model for this cervical cancer data. Moreover, this result is better than published work in [3].

Table 11: Comparison between published work in [3] and our result using the DT

|  | [3] result | Our result |
|---|---|---|
| **Sensitivity=TP** | 42.9 | 100 |
| **Precision** | 42.9 | 95 |
| **F-measure** | 30.6 | 97 |

# 5. Conclusion

To sum up, this article presents the comparison between different machine learning classifiers respect to the best predictive model for Cervical Cancer Dataset. Results show that this data is biased and addressing the imbalanced data is the first step for evaluation. Three techniques have been used to address the imbalanced data; over-sampling, under-sampling and combine both methods. Over-sampling yields better results than other two methods due to higher accuracy obtained by over-sampling . Further studies are conducted by using feature selection methods. Consequently, the SBS and SFS are superior techniques to enhance the performance of the prediction with accuracy 95%. By selecting the common features among the DT and KNN in Section 4.4, we obtain the best overall result with accuracy exceeds 97%. The selective features are Age, First sexual intercourse, number of pregnancies, Smokes, Hormonal Contraceptives and STDs:genital herpes. Interestingly, all six-selective feature make sense for diagnosing the cervical cancer. Eventually, we prefer DT classifer over GNB classifier. As a future work, selection features by LASSO method have not been tested with the classifiers to see their influence. Second potential work is multi-class classification with the four targets.